\newcommand{\slaninafiginline}[1]{}

\newcommand{\slaninafigure}[2]{
\begin{figure}[hb]
  \centering
  \vspace*{54mm}
  \includegraphics{#1.ps}
  \caption{#2}
  \label{fig:#1}
\end{figure}
}
\newcommand{\slaninafigurehh}[2]{
\begin{figure}[hb]
  \centering
  \vspace*{90mm}
  \includegraphics{#1.ps}
  \caption{#2}
  \label{fig:#1}
\end{figure}
}
\documentstyle[prl,aps,epsfig]{revtex}
\begin{document}
\draft

\twocolumn[\hsize\textwidth\columnwidth\hsize\csname@twocolumnfalse\endcsname

\title{Dynamical spin-glass-like behavior in an evolutionary game 
}

\author{Franti\v{s}ek Slanina}
\address{        Institute of Physics,
	Academy of Sciences of the Czech Republic,\\
	Na~Slovance~2, CZ-18221~Praha,
	Czech Republic\\
        e-mail: slanina@fzu.cz
}

\author{Yi-Cheng Zhang}
\address{Institut de Physique Th\'eorique, Universit\'e 
de Fribourg,
P\'erolles, CH-1700 Fribourg, Switzerland}
\maketitle
\begin{abstract}
We study a new evolutionary game, where players are tempted to take
part by the premium, but compete for being the
first who take a specific move. Those, who manage to escape the bulk
of players, are the winners.  While for large premium the game
is very similar to the Minority Game studied earlier, significant new
behavior, reminiscent of spin glasses is observed for premium below
certain level. 
 \end{abstract}

\twocolumn]

\section{Introduction}
The transfer of physical ideas and procedures to traditionally social
disciplines and a parallel backflow of inspiration for statistical
physics from these disciplines play an increasing role in the last
decade. 

Among other applications, the field of econophysics
\cite{ma_sta_99,bou_pot_00} attracted much attention. An impotrant
approach seems to be modelling the collective effect found in economic
systems by assemblies of individually acting heterogeneous agents. 
One of the basic models is
the Minority game \cite{cha_zha_97,cha_zha_98}, an evolutionary game
which mimics the adaptive behavior of agents with bounded
rationality, studied by W. B. Arthur in his El Farol bar problem
\cite{arthur_94}. 

In the Minority game, each player faces the choice between two
possibitities, which can be buying or selling stock, entering or not a
business, selecting first or second drive and so on.
The winning side is that of the minority of players. Each player
posesses a set of $S\ge 2$ strategies. Each strategy collects its
score, indicating the virtual gain of the player, if she had played
that strategy all the time. At each round, the playes choose among
their strategies the one with highest score. Such a sort of an on-line
learning and agaptation leads to better-than-random performance of the
system as a whole. It was found, that the properties of the game
depend on the memory length $M$ and number of players $N$ through the
scaling variable $\alpha=2^M/N$
\cite{cha_zha_98,sa_ma_ri_97,sa_ma_ri_99}. 
The Minoriry game was thoroughly studied both numerically and
analytically
\cite{jo_ha_hu_98,jo_le_hu_lo_99,ha_jef_joh_hu_00,ha_jef_joh_hu_00a,lo_hu_jo_00,ca_pla_gu_98,ca_pla_gui_99,li_ri_sa_99,li_ri_sa_99a,dhu_ro_99,dhu_rod_00,mansilla_99}
along with the study of the original bar attendance problem
\cite{jo_ja_jo_che_kwo_hui_98,jo_hu_zhe_tai_99} and various modifications of the Minority game
\cite{jo_hu_jo_lo_98,ceva_99,kal_schu_bri_00,pa_bas_99,dhu_ro_99a,sla_zha_99,slanina_00,ei_met_kan_kin_00}
An important role is attributed to the observation that the dynamics of the
memorized pattern and the strategies' scores are in certain regimes
decoupled \cite{cavagna_99,jo_hu_zhe_ha_99,cha_mar_00} and that the
thermal noise can ge introduced in the players' decisions 
\cite{ca_ga_gia_she_99,cha_mar_ze_00,gar_mor_sh_00}. 

The most intriguging feature is the minimum of
volatility, which occurs for the value $\alpha=\alpha_c\simeq
0.34$
\cite{cha_zha_98,sa_ma_ri_97,sa_ma_ri_99,jo_ja_jo_che_kwo_hui_98}. A
phase 
transition occurs here, the properties of which are well 
studied both numerically \cite{cha_ma_99} and analytically
\cite{cha_ma_ze_00,ma_cha_ze_99,cha_ma_zha_99,cha_ma_ze_99a,mar_cha_00}.  

The Minority game is essentially symmetric. The players can choose
between two sides, none of which is a priori preferred. The situation
is somewhat different in the bar attendance model 
\cite{jo_ja_jo_che_kwo_hui_98,jo_hu_zhe_tai_99}, where the optimal
attendance is set from outside. 
There are also variants of thge game, in which the players can decide
to participate or not, depending on their accumulated wealth
\cite{sla_zha_99,cha_ma_zha_99}. The number of players who influence
the outcome of the game can thus vary in time.

In this work we want to study a more abstract version of these models
with variable number of players. We implement a scheme, in which the
players struggle to be ``ahead'' of the other players, i. e. to come in
before the others and  not to stay if others have already left. 
This behavior is relevant in situations, where the early comers have
advantage, irrespectively of what is the direction of the movement. We
can think, for example, of a bunch of apes exploring a virgin
land. The first animal coming to a place finds enough food, but much
less is left for the followers. In an infinite space we could have
stationary movement of the bunch in one dorection. When the space
available is limited, frustration comes into play. The bunch should
oscillate between the borderlines and no-one can steadily win. 
Our aim is to formalize and simulate this situation in a similar
manner as the Minority game formalizes the inductive behavior.

\section{Escape game} 
We introduce a new mechanism, which leads to frustration in the
agent's actions. Similarly to the Minority game, we have $N$
agents. The agents can choose 
between two options: to participate (1) or not (0) in the
business. If we denote $a_j\in\{0,1\}$ the action of $j$'th player,
the attendance  is $A=\sum_j a_j$.  We measure the success of the
$j$'th player by her wealth $W_j$. The players who decide to
participate can be either rewarded or punished by 
corresponding change in their wealth, while
the wealth of non-participating agents remains unchanged.

There are two sources of the wealth change. First, there is a constant
influx of wealth into the system, which we will call premium $p$. All
participating players receive the amount $p/A$. Second, we reward the players,
who by some means induce the others to follow them in the next
step. If a player decided to participate in step $t-1$ and the
attendance rose from $t-1$ to $t$ (i. e. $A(t)-A(t-1)>0$), we consider
that the player was ``ahead'' of their companions and gets a
point. If, on the other hand the attendance decreased
(i. e. $A(t)-A(t-1)<0$), the player is considered as ``behind'' and
looses a point. If the attendance remains unchanged, no points are
assigned. Thus, each player tries to ``escape'' the bulk of the
other players. That is why we nicknamed the present dynamical multi-agent
system  ``Escape game''. 

As in the Minority game, the record is kept about the past $M$ changes of
the attendance, $\mu(t)=[c(t-1),c(t-2),...,c(t-M)]$. We denote $c=1$ increase
in $A$, and $c=-1$ decrease in $A$. There are two possibilities how to
deal with the case when the attendance does not change. It is possible
to attribute $c=0$ to such a situation; then the state variable will
have values from the set $c\in\{-1,0,+1\}$. Alternatively, we can merge
the cases of decrease and no change, so $c=-1$ also in the case when
the attendance is constant. In this case we distinguish only two
states, $c\in\{-1,+1\}$. We found, that both choices give
qualitatively similar results, while the former one leads to slightly more
demanding simulations. Therefore, throughout this article we we will
investigate the latter choice, with two states only.

The agents look at the record $\mu\in\{-1,+1\}^M$. They have a set of
$S$ strategies ($S=2$ in our simulations). The $s$-th strategy of the
$j$-th player prescribes for the record $\mu$ the action $a^\mu_{j,s}\in{0,1}$.
For each strategy, 
the score is  computed, which is the virtual gain 
of the player, if she played constantly that strategy. 
The update of the strategies' scores can be written as
\begin{eqnarray}
U_{j,s}(&&t+1)=U_{j,s}(t)+
\label{eq:update}\\
&&+a^{\mu(t-1)}_{j,s}\;\left(\frac{p}{A(t-1)}+
G (A(t)-A(t-1))\right)\;. \nonumber
\end{eqnarray}
For the function $G$, weighting the attendance changes, we use the
signum function, $G(x)={\rm sign}(x)$.
Note that the delivery of the player's gain is delayed: the action
taken at time $t-1$ can be rewarded anly at time $t$, so that it
influences the scores at time $t+1$. Again with close analogy to the
Minority game, the actions the players take are prescribed by the
strategies with highest score, $a_j=a_{j,s_{\rm M}}$,
where $s_{\rm M}$ denotes the most successfull strategy at the moment, 
$U_{j,s_{\rm M}}=\max_s U_{j,s}$. 

To see clearly the points of difference from the Minority game, note
first that in the Escape game the non-participation $a_j=0$ cannt change the
wealth of the player. Second, the strategie's update rule
(\ref{eq:update}) contains derivative of the attendance, not the
attendance itself.

\section{Evolution of attendance and glassy behavior}

Let us see qualitatively first, what is the time dependence of the attendance.
We have found that for large enough premium the Escape game behaves
in very similar manner to the Minority game. The attendance
fluctuations decrease from its initial value until they stabilize at a
stationary value. The average attendance is shifted from its random
value $N/2$ above, as a response to the incentive, posed by the premium,
for the payers to prefer presence over absence.
We can see in 
Fig. \ref{fig:aws-vs-t-1}
an example of such a behavior. The stationary state is reached in a
short time (shorter than $10^5$ steps for $N=200$). The wealth of the
most successful player grows constantly at high rate and also the
average wealth slightly grows. This means, that due to the premium the
game is a positive-sum game on average. 

\slaninafigurehh{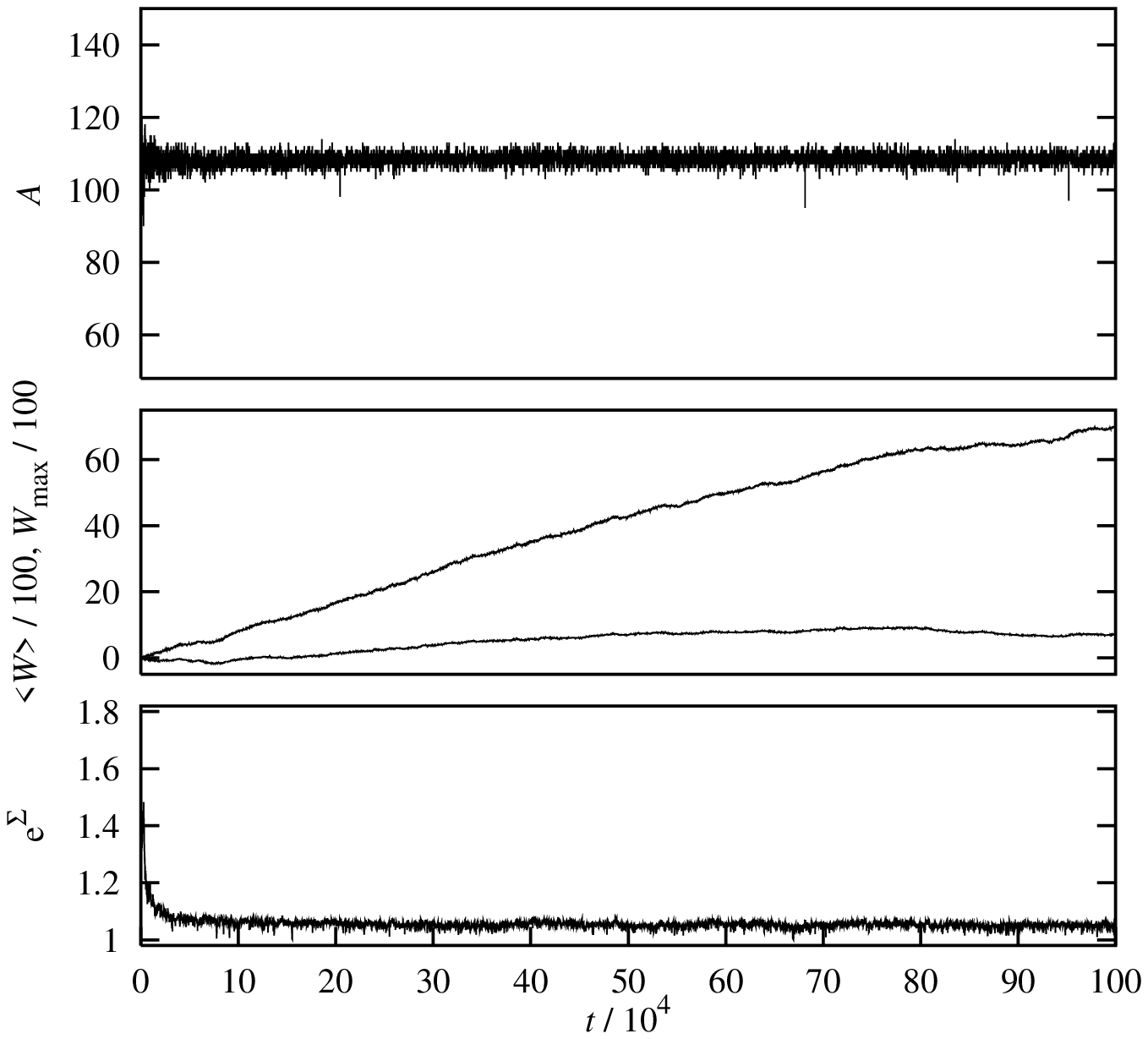}{
Example for the time dependence of the attendance (top frame), the
effective number of 
strategies (bottom frame), and average wealth (middle frame, lower
curve) and maximum wealth among all players (middle frame, upper curve).
The number of players is $N=200$, memory $M=5$ and premium $p=1$.
}

We can also observe the time
dependence of the effective number of strategies per player ${\rm
e}^\Sigma$. We 
define it through the average entropy of the usage of the strategies,
\begin{equation}
\Sigma(t) = \frac{1}{N}\sum_{j,s} \nu_{j,s}(t)\log\nu_{j,s}(t)
\end{equation}
where $\nu_{j,s}$ is the time-averaged frequency of the usage of
$s$-th strategy of $j$-th player. We performed the time averaging over
a relatively short window using the exponential weighting,
$\nu_{j,s}(t)=(1-\lambda)u_{j,s}(t)+\lambda \nu_{j,s}(t-1)$, where the
usage index $u_{j,s}(t)=1$ if the player $j$ used the strategy $s$ at
time $t$, and $u_{j,s}(t)=0$ otherwise. We used $\lambda=0.99$, which
corresponds to effective time-window width 100 steps.

We can see in Fig. \ref{fig:aws-vs-t-1} that the effective number of used strategies stabilizes at a value above but
close to 1, which also corresponds well to the behavior of the
Minority game in the symmetry-broken phase.

\slaninafigurehh{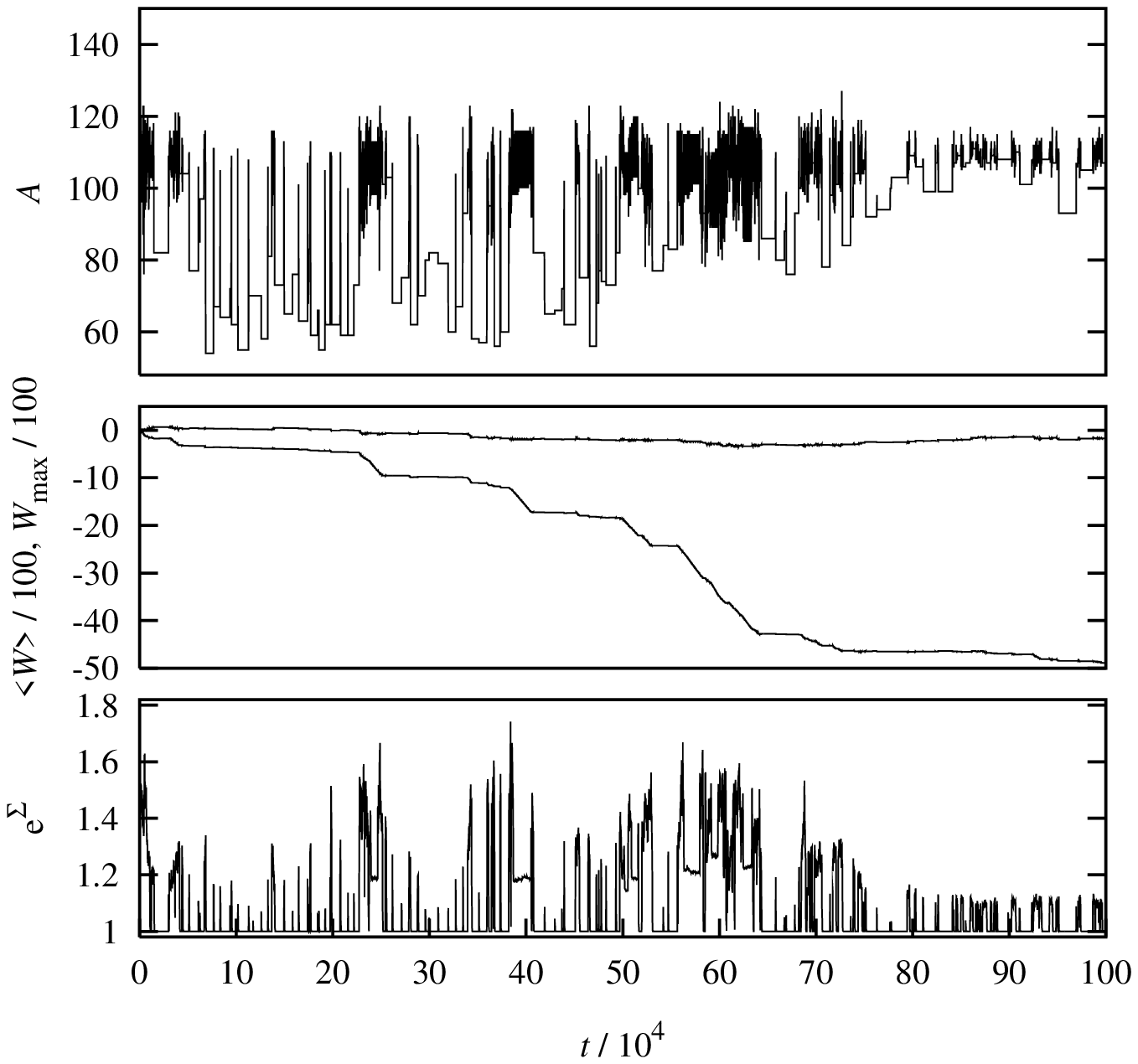}{The same as in
Fig. \ref{fig:aws-vs-t-1}, but for premium $p=0.01$.
}

On the other hand, we can observe significant change of the behavior
of the model, if the premium is decreased below certain level.
An example can be seen in Fig. \ref{fig:aws-vs-t-0.01a}.
First, the transient time before the system settles in a stationary
state is significantly larger (we observed, that it is nearly $10^6$
steps). Then, we can see that periods with significant attendance
fluctuations are alternating with periods with constant attendance,
which can frequently last more than $10^4$ steps. The fluctuating
periods are characterized by decrease of the average wealth (the game
is a negative-sum one) and effective number of strategies larger than
1. On the other hand, the periods of constant attendance exhibit
nearly constant average wealth and effective number of strategies
equal to 1. It means, that within the constant periods the system is
able to find a favorable configuration, in which the game is globally
effective (avoids the losses due to fluctuations) and everybody is
stuck at single strategy. Moreover, there are many of such states: in
different constant periods the attendance may differ considerably.

\slaninafigurehh{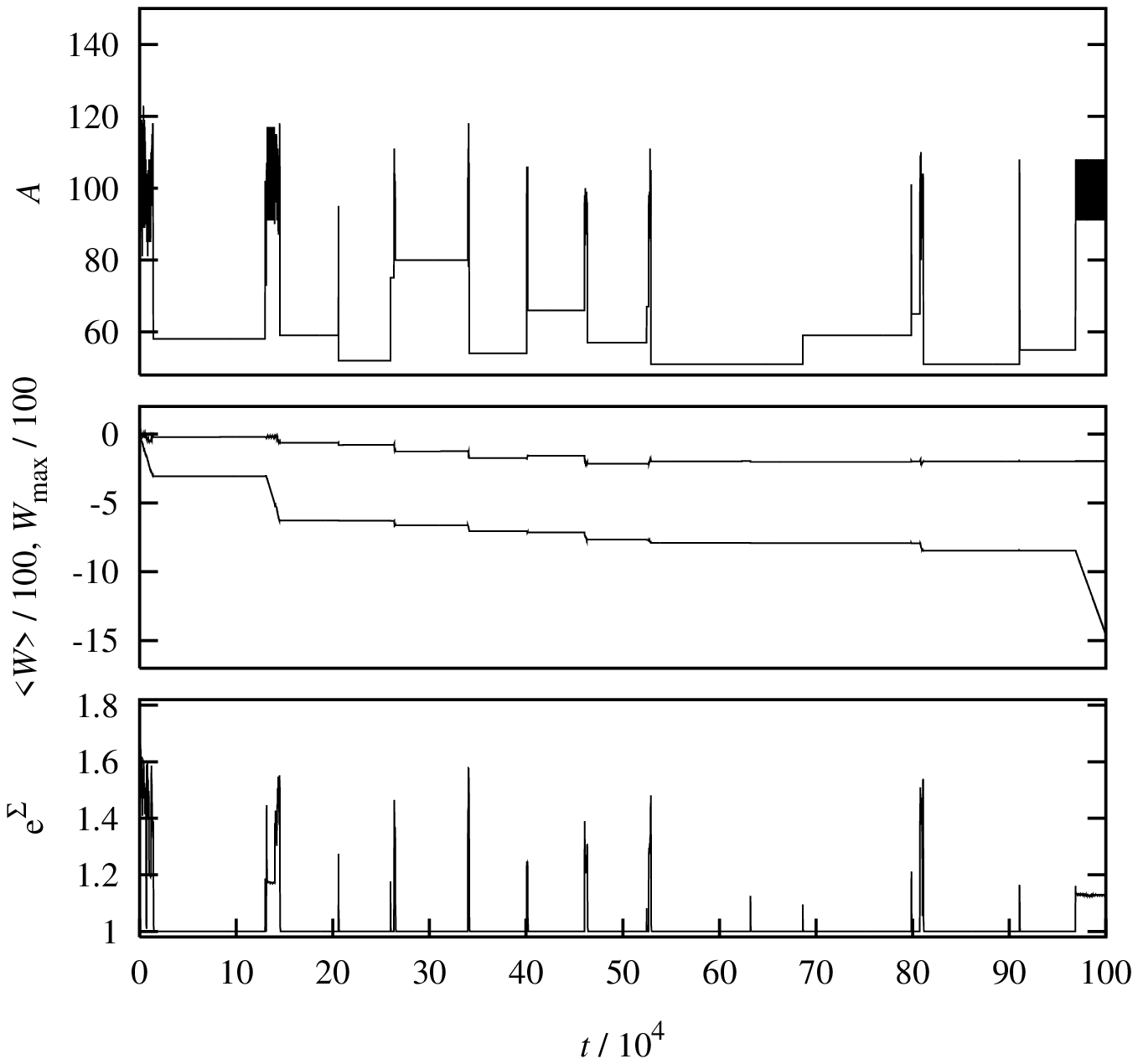}{The same as in Fig. \ref{fig:aws-vs-t-1}, but
for premium $p=0.001$.
}

This behavior is further pronounced when we diminish the premium
even more. We can see such a behavior in
Fig. \ref{fig:aws-vs-t-0.001b}.
We can observe long periods (sometimes longer than $10^5$ steps) of
constant attendance, separated by very short fluctuating periods. The
effective number of strategies differs from 1 only during these short
periods. Again, there are many configurations in which the attendance
does not change for long time.

The situation is reminiscent of spin glasses. In the spin glass
behavior, there are many states, stable for long time, but mutually
very different, in which the system can stay. If the bariers between
these states are not infinite (which happens only in the fully
connected case in thermodynamic limit), the dynamics of such a system
is very similar to our Escape game: long periods of stasis within one
state, interrupted by short periods coresponding to the jumps from one
state to another.

Therefore, we can describe the behavior with changing premium $p$ as a
kind of transition from ``paramagnet'' (high $p$) to ``spin-glass''
phase (low $p$). We have not studied in detail the phase
diagram. However, we observed, that with fixed $N$ the transition
occurs at smaller $p$ if the memory is longer.

\section{Time-averaged attendance and its fluctuations}

The response of the player's assembly to the premium was measured by
the average attendance. Its dependence on the value of premium is
shown in
Fig. \ref{fig:att-vs-p-7and5}. As expected, it is an increasing
function. We can see, that shorter memory leads to more strong
response to the premium. The value of $p$ at which the average
attendance crosses its random value $N/2$ is smaller for shorter
memory.

\slaninafigure{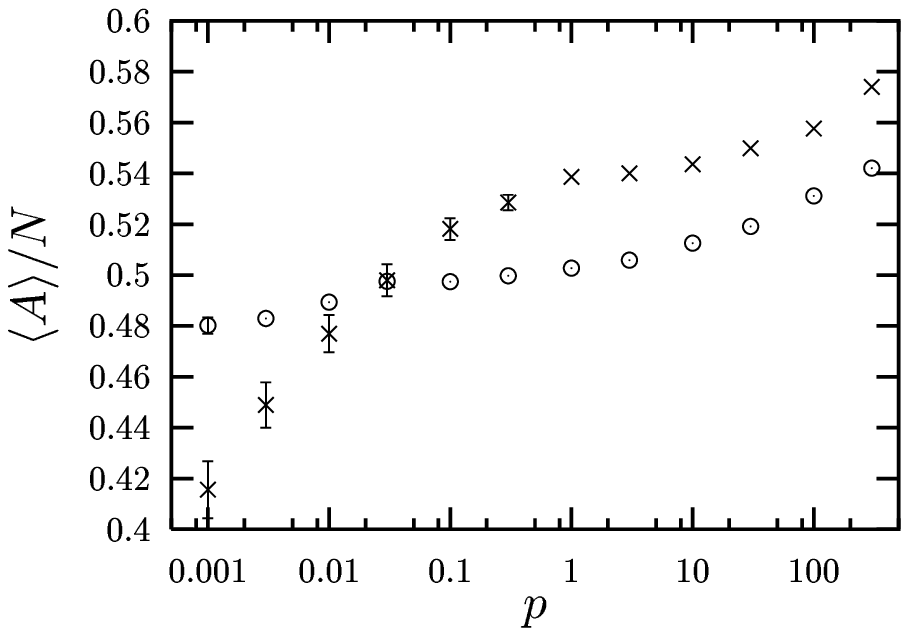}{Average attendance for $N=200$ and
memory length $M=5$ ($\times$) and $M=7$ ($\circ$).  The data are
averaged over $50$ independent runs. Each run was $10^6$
steps long and the average was taken over $5\cdot 10^5$ last steps. 
Where not shown, the error bars are smaller than symbol size.}
\slaninafigure{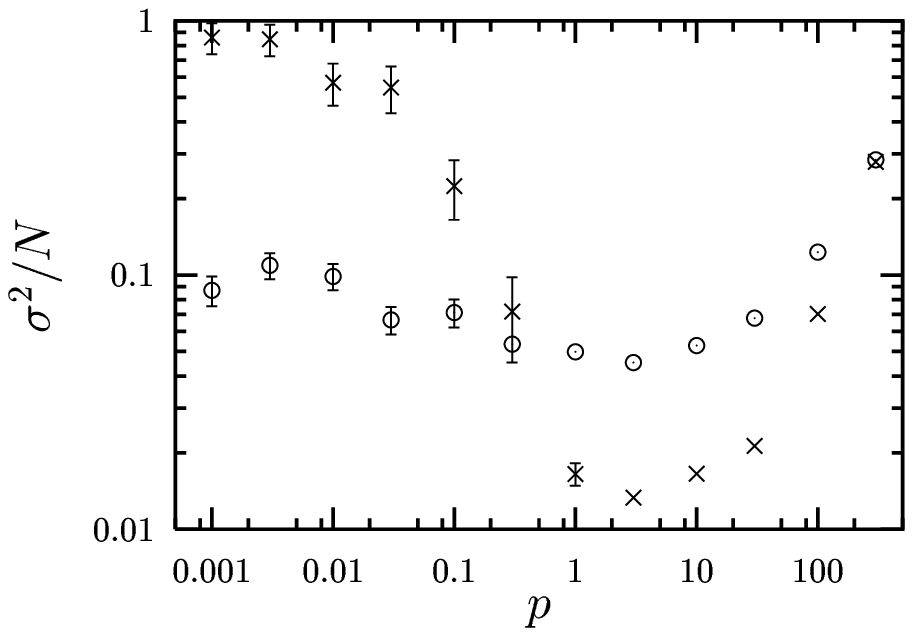}{Attendance fluctuations for $N=200$ and
memory length $M=5$ ($\times$) and $M=7$ ($\circ$).  The data are
averaged over $50$ independent runs. Each run was $10^6$
steps long and the average was taken over $5\cdot 10^5$ last steps. 
Where not shown, the error bars are smaller than symbol size.}

The dependence of the attendance fluctuations $\sigma^2$ on $p$ is shown in
Fig. \ref{fig:sig-vs-p-7and5}. Again, the dependence on $p$ is more
pronounced for shorter memory. The minimum, which is located around
$p=3$ for $N=200$ is probably connected with the transition from the
``paramagnet'' to ``spin-glass'' phase: for low $p$ the fluctuations
are mainly due to rare but large jumps of the attendance from one
quasi-static value to another.
Indeed, we observed qualitatively that the transition occurs somewhat
below the position of the minimum in $\sigma^2$.

\slaninafigure{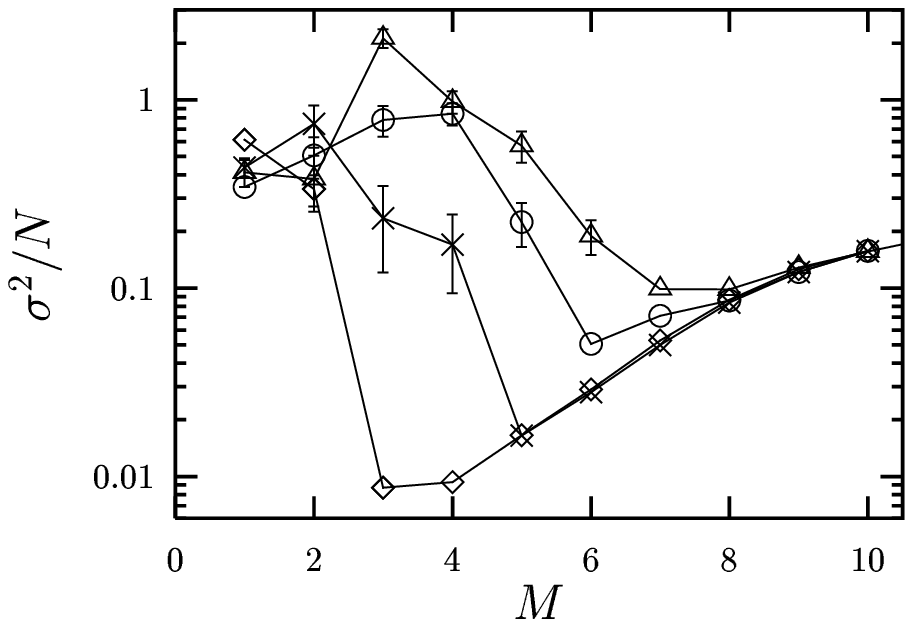}{
Average attendance fluctuations for $N=200$ and premium $p=10$ ($\Diamond$), 1
($\times$), 0.1 ($\circ$), and 0.01 ($\triangle$). The data are
averaged over $50$ independent runs. Each run was $10^6$
steps long and the average was taken over $5\cdot 10^5$ last steps. 
Where not shown, the error bars are smaller than symbol size.
}
\slaninafigure{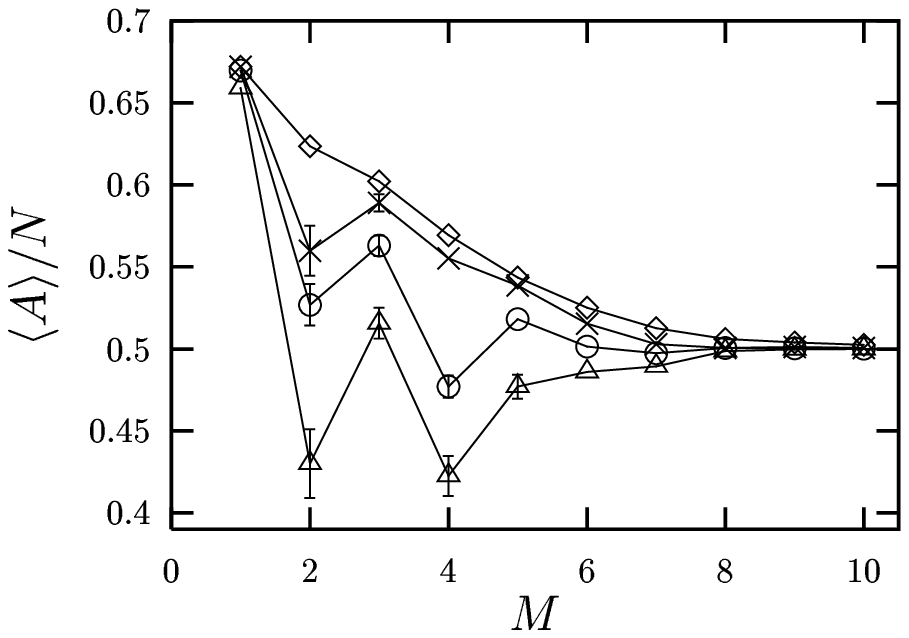}{
Average attendance for $N=200$ and premium $p=10$ ($\Diamond$), 1
($\times$), 0.1 ($\circ$), and 0.01 ($\triangle$). The data are
averaged over $50$ independent runs. Each run was  $10^6$ steps long
and the average was taken over $5\cdot 10^5$ last steps 
steps. Where not shown, the error bars are smaller than symbol size.
}

The memory dependence of attendence fluctuations for several valuer of
$p$ is presented in
Fig. \ref{fig:sig-200-all}. In the minimum of fluctuations we
recognize the same behavior as in the Minority game. Here, however,
the position of the minimum depends strongly on the value of the
premium. We can see, that the long-memory phase does not depend much
on the premium. On the other hand, the crowded phase is strongly
influenced by the premium. For smaller $p$ the crowded phase occurs at
longer memories. 

In Fig. \ref{fig:att-200-all}
the dependence of average attendance on $M$ is shown. 
We can clearly observe that, as discussed already with
Fig. \ref{fig:att-vs-p-7and5}, longer memory supresses the response of
the system to the premium. For shorter memories and sufficiently small
$p$ (for $N=200$ it means $M\le 4$ and $p\le 0.1$), we observe an
interesting, yet not clearly understood behavior, characterized by
non-monotonic dependence of the  average attendance on $M$.

\slaninafigure{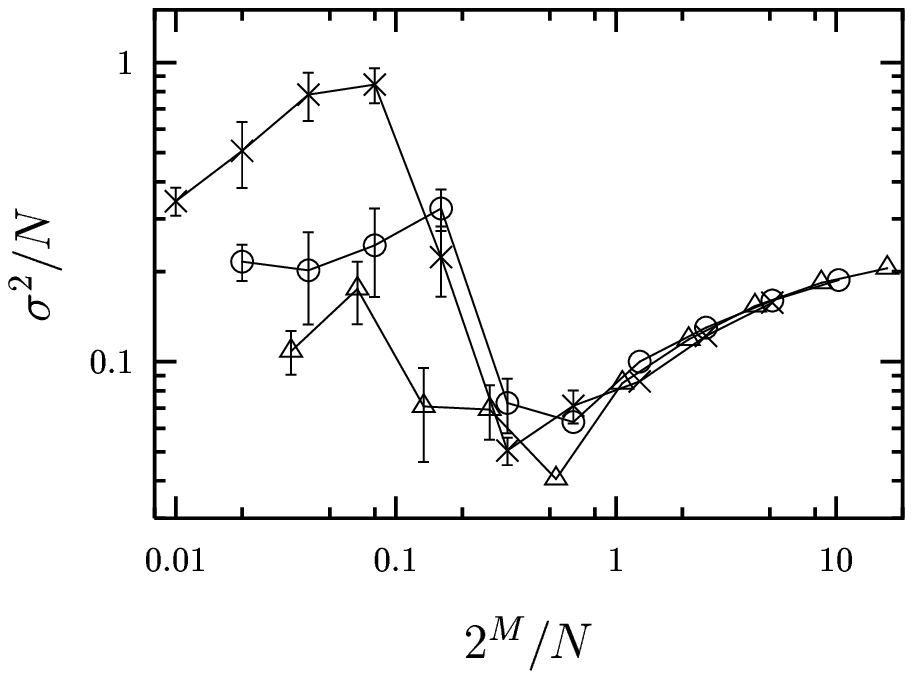}{
Rescaled attendance fluctuations for premium $p=0.1$ and number of
players $N=200$ ($\times$), 100 ($\circ$), and 60 ($\triangle$).  The data are
averaged over $50$ independent runs. Each run was  $10^6$ steps long
and the average was taken over $5\cdot 10^5$ last steps 
steps. Where not shown, the error bars are smaller than symbol size.
}
\slaninafigure{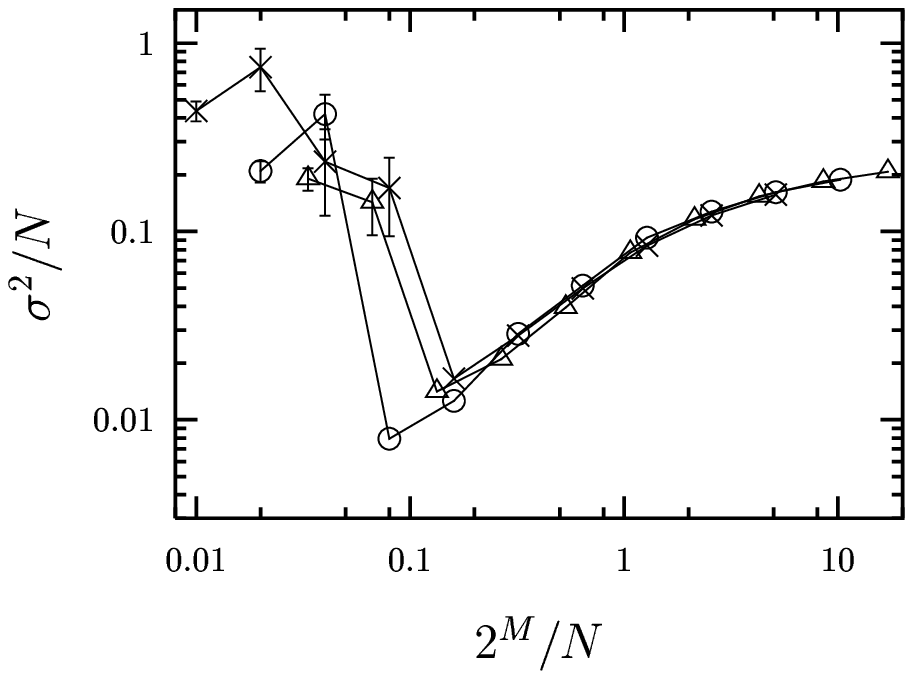}{
The same as in Fig. \ref{fig:sig-scaled-0.1}, but for premium $p=1$.
}

In the minority game the relevant quantities are functions of the
scaling variable $2^N/N$. We tried the same scaling also in our Escape
game.
The results for $p=0.1$ are in 
Fig. \ref{fig:sig-scaled-0.1}, while 
the case $p=1$ is shown in
Fig. \ref{fig:sig-scaled-1}.
As we already noted, larger $p$ makes the game be behave more
similarly to usual Minority game. Here it is illustrated by the fact,
that the data collapse is much better for $p=1$ than for $p=0.1$. We
can also see again that the in large-memory phase the scaling works
very well, while the crowded phase behaves differently.

\section{Conclusions}

We introduced an evolutionary game, which we called Escape game. The
dynamical rules are similar in priciple to the Minority game, but some
substantial differencs occur. 
We observed that in the large-memory phase the response to the premium
is lower. At the same time, the behavior is closer to the Minority
game, which is quantitatively seen in the fact, that the relevant
quantities depend on the scaling variable $2^M/N$ as in the Minority
game. On the other hand, for fixed memory length the behavior is
similar for large premium but different from it for small premium. 

The fact that the players are drawn
into the play by the premium leads to enhanced attendance for high
premium. On the other hand, for small premium the incentive is not
strong enough and the system starts to exhibit a new dynamical phase,
whose character is very close to a spin-glass.

This result can be qualitatively understood as follows. The presence
of fluctuations makes the game a negative-sum one, if we neglected the
premium. High premium overweights the negative efect of fluctuations
and the players choose the strategies, which give them highest
possible attendance, irrespectively of fluctuations they cause. That
is why for higher premium both the attendance and its fluctuations
grow.

For small premium the fluctuations are disastrous and the players tend
to avoid them by not participating. The situations, in which there are
two groups, one of constantly participating players and second of
constantly non-patricipating ones are suitable for both groups. The
first one receives steadily the small premium, the second one at least
does not loose anything. After some time, however, the scores of the
absent players change so that they would prefer being in over staying
out. At that moment, the system is reshuffled and another
configuration of participating and non-participating players is
found. The mean period of such reshufflings must be longer for smaller
premium, because in this case the scores change more slowly. 
We expect, that the transition occurs at such a value of the premium,
which would correspond to reshuffling period of order one.
So, the ``spin-glass'' behavior is purely dynamical in
origin. 

To sum it up, we
can draw a fuzzy line in the $M$ versus $p$ plane, which encloses the
region of both $p$ and $M$ small. Outside this region the Escape game 
behaves merely as a sligt modification of the usual Minorit game, while
inside we observe qualitatively new features, including the dynamical
``spin-glass'' behavior.

\acknowledgments{We woulkd like to express our gratitude to P. De Los Rios and
D. Challet for useful discussions and comments.   
}

\end{document}